\documentclass[useAMS,usenatbib]{mn2e}

\usepackage{epsfig,amsmath,amssymb,amsfonts,mathrsfs,latexsym,graphicx,lscape}
\usepackage{times}
\usepackage{longtable}
\usepackage{xcolor}
\input{epsf}
\usepackage{multirow}
\usepackage{natbib}
\usepackage{makecell}
\usepackage{color}

\title[Timing view of heartbeat state]
{A timing view of the heartbeat state of GRS 1915+105}

\author[Yan et al.]
       {Shu-Ping Yan$^{1,2,6}$\thanks{E-mail: yanshuping@pmo.ac.cn}, 
        Li Ji$^{1,2}$,
        Mariano M\'endez$^{3}$,
        Si-Ming Liu$^{1,2}$,
        Na Wang$^{4}$,
\newauthor 
        Xiang-Dong Li$^{5,6}$,
        Ming-Yu Ge$^{7}$,
        Jin-Yuan Liao$^{7}$,
        Shu Niu$^{1,2}$,
        Jin-Lu Qu$^{7}$,
\newauthor 
        Guo-Qiang Ding$^{4}$,
        Qing-Zhong Liu$^{1}$,
        Wei Sun$^{1,2}$\\
      $^{1}$Purple Mountain Observatory, Chinese Academy of Sciences, Nanjing 210008, China\\
      $^{2}$Key Laboratory of Dark Matter and Space Astronomy, Chinese Academy of Sciences, Nanjing 210008, China\\
      $^{3}$Kapteyn Astronomical Institute, University of Groningen, P.O. Box 800, 9700 AV Groningen, The Netherlands\\
      $^{4}$Xinjiang Astronomical Observatory, Chinese Academy of Sciences, Xinjiang 830011, China\\
      $^{5}$Department of Astronomy, Nanjing University, Nanjing 210093, China;\\
      $^{6}$Key Laboratory of Modern Astronomy and Astrophysics (Nanjing University), Ministry of Education, Nanjing 210093, China\\
      $^{7}$Key Laboratory for Particle Astrophysics, Institute of High Energy Physics, Chinese Academy of Sciences, Beijing 100049, China}

\begin{document}

\date{Accepted . Received ; in original form }

\pagerange{\pageref{firstpage}--\pageref{lastpage}} \pubyear{2015}

\maketitle

\label{firstpage}

\begin{abstract}

We present a timing analysis of two {\it Rossi X-ray Timing Explorer} observations of the microquasar GRS~1915+105 during the heartbeat state. The phase-frequency-power maps show that the intermediate-frequency aperiodic X-ray variability weakens as the source softens in the slow rise phase, and when the quasi-periodic oscillation disappears in the rise phase of the pulse of the double-peaked class its sub-harmonic is still present with a hard phase lag. In the slow rise phase, the energy-frequency-power maps show that most of the aperiodic variability is produced in the corona, and may also induce the aperiodic variability observed at low energies from an accretion disk, which is further supported by the soft phase lag especially in the intermediate-frequency range (with a time delay up to 20 ms). In the rise phase of the pulse, the low-frequency aperiodic variability is enhanced significantly and there is a prominent hard lag (with a time delay up to 50 ms), indicating that the variability is induced by extension of the disk toward small radii as implied by the increase in flux and propagates into the corona. However, during the hard pulse of the double-peaked class, the variability shows no significant lag, which may be attributed to an optically thick corona. These timing results are generally consistent with the spectral results presented by \citet{Neilsen11,Neilsen12} which indicated that the slow rise phase corresponds to a local Eddington limit and the rise phase of the pulse corresponds to a radiation pressure instability in the disk.
\end{abstract}


\section{INTRODUCTION}

Accretion plays a crucial role in the evolution of black hole binaries (BHBs). Timing analysis is an important tool for studying the accretion flow. The Fast Fourier Transformation is one of the most popular methods of timing analysis. However, the origins of the X-ray aperiodic variabilities and the low-frequency quasi-periodic oscillations (LFQPOs) in Fourier power density spectra (PDS) from BHB accretion flows remain yet unsolved \citep[e.g.][]{Miller14b}. 

GRS~1915+105 is a BHB suitable for studying the X-ray variability. The source has a rapidly spinning black hole \citep{Zhang97, McClintock06, Middleton06, Miller13} with a mass of $\sim 12$ $M_{\odot}$, and a K-M III secondary with a mass of $\sim 0.8$ $M_{\odot}$ \citep{Reid14, Harlaftis04, Greiner01b}. It is $\sim 10$ kpc away from the earth \citep[e.g.][]{Fender99, Zdziarski05, Reid14}, and is considered a microquasar because it shows a relativistic jet whose inclination to the line of sight is $\gtrsim 60^\circ$ \citep{Mirabel94, Fender99, Reid14}. GRS~1915+105 has been active for more than twenty years \citep{Castro92}, and is intensively observed with the \emph{Rossi X-ray Timing Explorer} ({\it RXTE}). The various kinds of source variabilities \citep{Belloni00, Klein02, Hannikainen05} and the abundance of LFQPOs \citep[e.g.][]{Morgan97, Chen97, Strohmayer01, Belloni01, Belloni06, Yan13mn} displayed in the {\it RXTE} data make it a unique source for studying the X-ray variability.

The variabilities of GRS~1915+105 are classified into 14 classes based on the count rate and the colour characteristics of the source \citep{Belloni00, Klein02, Hannikainen05}. Each class is regarded as transitions among three states, A, B and C. State C is a low-luminosity, spectrally hard state while states A and B are high-luminosity, soft states. The $\rho$ class (heartbeat state) is a peculiar class where the source oscillates quasi-periodically between states B and C. The light curves in the $\rho$ class have one to several peaks per $\rho$-cycle \citep[e.g.][]{Taam97, Vilhu98, Belloni00, Massaro10}. We call the $\rho$ class with one peak per cycle the single-peaked $\rho$ class (hereafter called $\rho_1$ class), and the $\rho$ class with two peaks per cycle the double-peaked $\rho$ class (hereafter called $\rho_2$ class). 

\citet{Neilsen11,Neilsen12} carried out a phase-resolved spectral analysis of the {\it RXTE} observations 60405-01-02-00 during the $\rho_2$ class and 40703-01-07-00 during the $\rho_1$ class. In order to investigate the origin of the LFQPO, \citet{Yan13apj} performed a phase-resolved timing analysis of the $\rho_2$ class, and showed that the LFQPO was tightly related to the corona. Besides, for one phase interval of the $\rho_2$ class \citet{Yan13apj} first detected a high-frequency ($\gtrsim$ 10 Hz) aperiodic variability from the disk through the amplitude-ratio spectrum method. 

In this paper we present the results of the diagram of the power density as a function of Fourier frequency and $\rho$-cycle phase (phase-frequency-power map), the diagram of the power density as a function of Fourier frequency and photon energy (energy-frequency-power map), and the phase lag as a function of Fourier frequency (phase-lag spectrum) of the $\rho_1$ and $\rho_2$ classes. We use the phase-frequency-power maps to investigate the continuous evolution of the X-ray variability along with the spectral evolution, use the energy-frequency-power maps to visually display the correlation between the X-ray variability and the spectral components, and use the phase-lag spectra to obtain the sequence information of the X-ray variabilities from different spectral components.

In addition to studying the origin of the X-ray variability, we aim to acquire a spectral-timing unified picture of the accretion in the heartbeat state of GRS 1915+105 through a combination analysis of our timing results and the spectral results presented by \citet{Neilsen11,Neilsen12}.

We describe the observations and the data reduction methods in Section \ref{sec:data}, present the results in Section \ref{sec:result}, show the discussion in Section \ref{sec:discuss}, and list the conclusions in Section \ref{sec:con}.

\section{OBSERVATIONS AND DATA REDUCTION}\label{sec:data}

Using the phase-folding method of \citet{Neilsen11}, \citet{Yan13apj} performed a phase-resolved timing analysis for the $\rho_2$ class ({\it RXTE} observation 60405-01-02-00 on 2001 May 23 at 11:18:42 UT with 13.9 ks exposure time). In this paper with the same method we carry out a timing analysis for the $\rho_1$ class ({\it RXTE} observation 40703-01-07-00 on 1999 February 26 at 07:32:13 UT with 9.9 ks exposure time). We extract two dead-time-corrected and background subtracted light curves with a time resolution of 1 s from the binned-mode data (B\_8ms\_16A\_0\_35\_H\_4P) in the 1.9--13.0 keV band and the event-mode data (E\_16us\_16B\_36\_1s) in the 13.0--60 keV band respectively, and add them together to obtain a light curve. We apply the barycenter correction to the light curve and then fold it to obtain an average folded light curve. We determine the start time of each cycle by an iterative cross correlation method \citep[for more details see][]{Neilsen11}, and obtain 209 cycles with a mean period of 44.54 s for the $\rho_1$ class. In \citet{Yan13apj} we obtained 257 cycles with a mean period of 50.29 s for the $\rho_2$ class. We average the PDS from a certain phase interval of all cycles for each class to study the timing properties of the $\rho$ oscillation, since the shape difference of the individual cycles is always less than 20\% for a given phase interval.

We calculate the PDS in different energy bands from the binned and event files using 2 s segments at a time resolution of 8 ms, and subtract the dead-time-corrected Poisson noise (Morgan et al. 1997) from each PDS and normalize them to units of (RMS/mean)$^2$/Hz \citep[e.g.][]{Miyamoto92}. We then fit the PDS with a model including several Lorentzians to represent the broad-band noise and the LFQPOs \citep{Nowak00, Belloni02}. We calculate the PDS in the 1.9--60 keV band for each 0.04 phase interval of the $\rho_1$ class, and calculate the PDS in the 2.1--60 keV band for each 0.02 phase interval of the $\rho_2$ class. We then use these PDS to produce the phase-frequency-power maps for the $\rho_1$ and $\rho_2$ classes. We compute the significance of the QPOs with the method adopted by \citet{Strohmayer2005}. The 3.9 Hz QPO in the phase 0.84--0.86 of the $\rho_2$ class has the minimum significance of 0.0017 among all of the QPOs studied.

We produce the PDS in different energy bands for phases I (0.00--0.08), II (0.08--0.26), III (0.26--0.40), IV (0.40--0.60), and V (0.60--1.00) of the $\rho_1$ class, and for phases i (0.02--0.12), ii (0.12--0.26), iii (0.26--0.40), iv (0.40--0.74), v (0.74--0.92), and vi (0.92--0.02) of the $\rho_2$ class, and use these PDS to produce the energy-frequency-power maps. We correct these PDS for background due to the energy dependence of the background \citep{Berger94, Rodriguez11}.

For these phases we calculate the Fourier cross power spectra of two light curves extracted in different energy bands following \citet{Nowak99} to obtain the phase-lag spectra. For the $\rho_1$ class, a positive phase lag denotes that the variability in the 5.1--38.4 keV band lags that in the 1.9--5.1 keV band. For the $\rho_2$ class, a positive phase lag denotes that the variability in the 4.9--37.8 keV band lags that in the 2.1--4.9 keV band. The reference bands for the phase lag calculations for the two classes cannot be the same due to that they were observed in different gain epochs while archived with the same binning modes. All error bars in this paper correspond to 1$\sigma$ confidence level.

\begin{figure}
\centerline{\includegraphics[height=8.5cm,angle=-90]{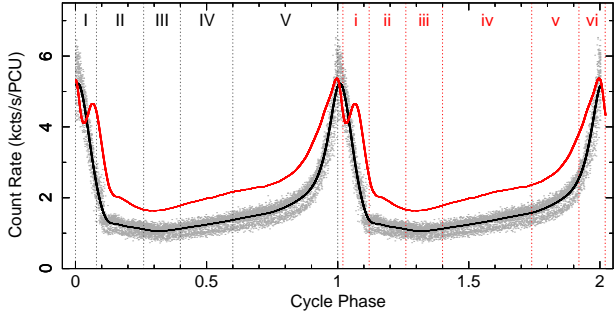}}
\caption{\label{fig:lan.lc} The phase-folded PCA light curves of the $\rho_1$ class (black; \emph{RXTE} Observation 40703-01-07-00) and the $\rho_2$ class (red; \emph{RXTE} Observation 60405-01-02-00) in GRS 1915+105. The gray points are the data points of the $\rho_1$ class. The black and red vertical dashed lines denote phases I (0.00--0.08), II (0.08--0.26), III (0.26--0.40), IV (0.40--0.60), V (0.60--1.00) of the $\rho_1$ class, and phases i (0.02--0.12), ii (0.12--0.26), iii (0.26--0.40), iv (0.40--0.74), v (0.74--0.92), vi (0.92--0.02) of the $\rho_2$ class.}
\end{figure}

\begin{figure}
\centerline{\includegraphics[height=12.5cm,angle=0]{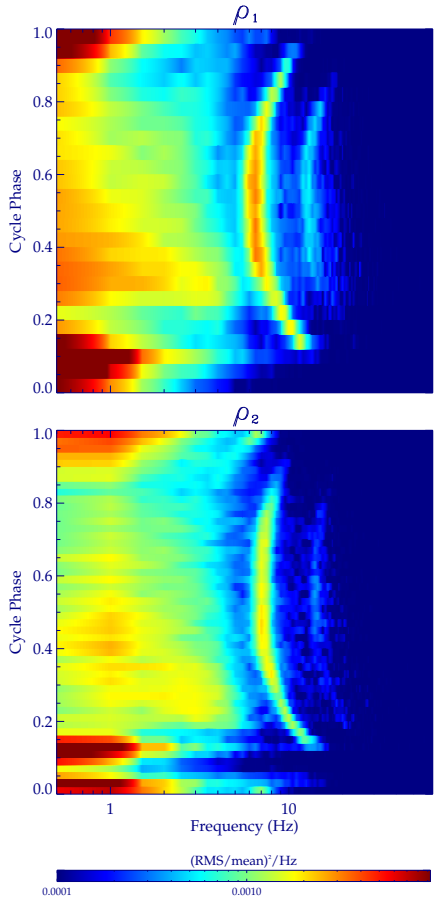}}
\caption{\label{fig:f2t} The phase-frequency-power maps of the $\rho_1$ class (upper panel) and the $\rho_2$ class (lower panel) in GRS 1915+105. The colour bar shows the scale of power density in the plots. The power densities lower than $10^{-4}$ (RMS/mean)$^2$/Hz are set to blue.}
\end{figure}

\begin{figure}
\centerline{\includegraphics[height=8.3cm,angle=-90]{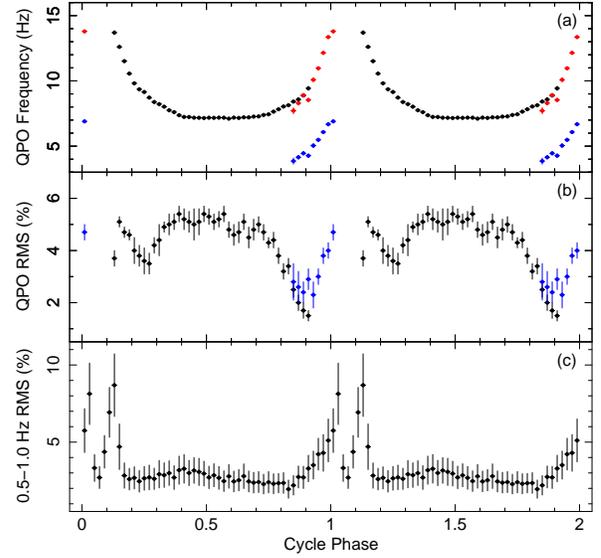}}
\caption{\label{fig:Rho2QPO} The evolutions of the QPO and its sub-harmonic as well as the 0.5--1.0 Hz aperiodic variability of the $\rho_2$ class in GRS 1915+105. In panels (a) and (b), the black points represent the fundamental QPO in phase 0.12--0.92, the blue points represent the sub-harmonic in phase 0.84--0.02. We doubled the frequencies of the sub-harmonic and show these values as red points.}
\end{figure}

\section{RESULTS}\label{sec:result}

We present in this section the phase-frequency-power maps, the energy-frequency-power maps, and the phase-lag spectra of the $\rho_1$ and $\rho_2$ classes. We show the phase-folded {\it RXTE}/Proportional Counter Array (PCA) light curves and the $\rho$-cycle phase divisions of the two $\rho$ classes in Figure \ref{fig:lan.lc}. For the aperiodic variability we define three frequency bands, the low-frequency band ($\lesssim$ 1 Hz), the intermediate-frequency band ($\sim$ 1--10 Hz), and the high-frequency band ($\gtrsim$ 10 Hz).

\begin{figure}
\centerline{\includegraphics[height=17.5cm,angle=0]{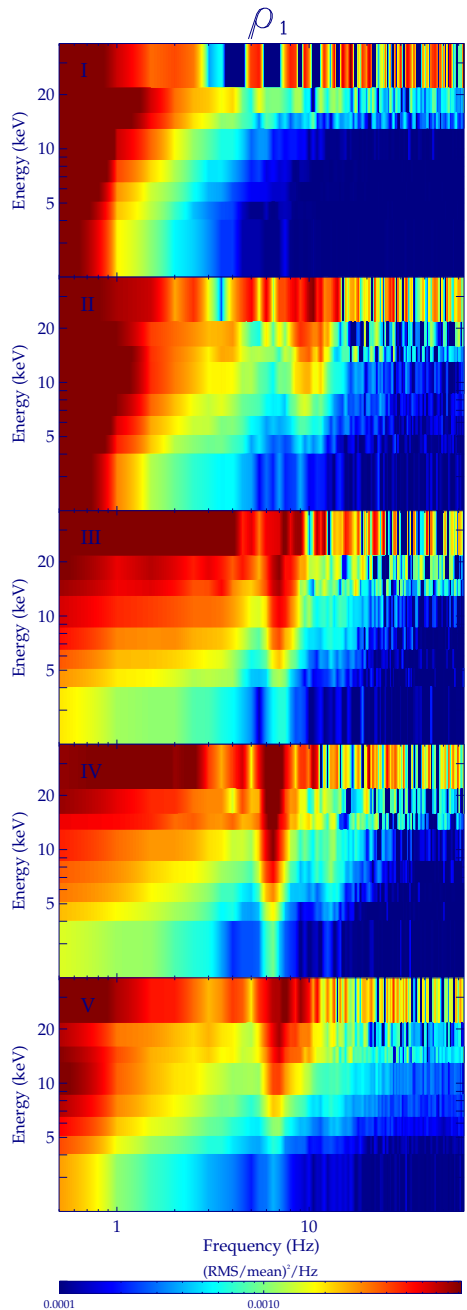}}
\caption{\label{fig:F2E40703} The energy-frequency-power maps for different phases of the $\rho_1$ class in GRS 1915+105. The colour bar is the same as that in Figure \ref{fig:f2t}.}
\end{figure}

\begin{figure}
\centerline{\includegraphics[height=21cm,angle=0]{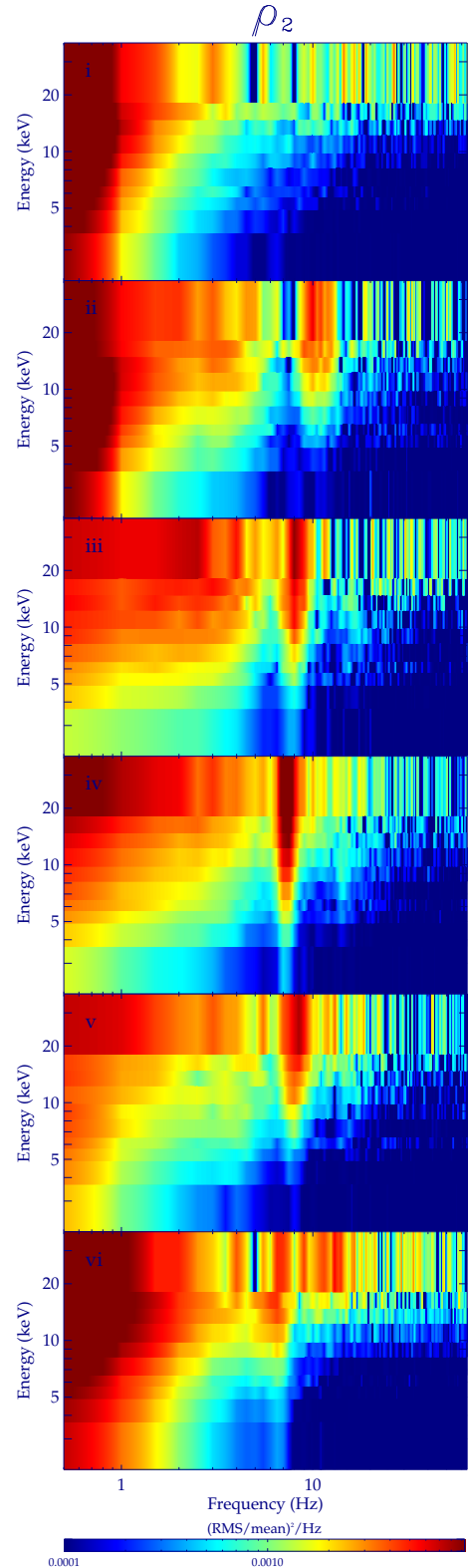}}
\caption{\label{fig:F2E60405} Same as Figure \ref{fig:F2E40703} for different phases of the $\rho_2$ class in GRS 1915+105.}
\end{figure}

\subsection{Phase-frequency-power maps}

Figure \ref{fig:f2t} shows the phase-frequency-power maps. The colours denote the values of the power density. The power densities lower than $10^{-4}$ (RMS/mean)$^2$/Hz are set to blue. 

For the $\rho_1$ class the evolutions of the LFQPO and its second harmonic are visible as a green and a cyan fringe, respectively. The red, yellow, green, cyan, and blue regions, which are clearly separated in the maps, show the evolution of the aperiodic variability. As the $\rho$-cycle phase, $\phi$, increases from $\sim 0.25$ to $\sim 0.80$, the transition frequency between the red and green regions, $f_{\rm tr1}$, decreases from $\sim 3.5$ Hz to $\sim 0.7$ Hz, the transition frequency between the green and blue regions, $f_{\rm tr2}$, decreases from $\sim 5.0$ Hz to $\sim 3.0$ Hz. At $\phi \sim 0.20$ and $\sim 0.85$, which correspond to the end and the start of the flux peak respectively, $f_{\rm tr2}$ decreases to $\sim 2.0$ Hz and $\sim 1.2$ Hz respectively. At $\phi \sim 0.20$, $f_{\rm tr1}$ decreases to $\sim 0.5$ Hz. In the phase $\sim$ 0.92--0.15 the power density of the aperiodic variability below $\sim 1.5$ Hz is enhanced.

For the $\rho_2$ class the evolutions of the LFQPO and its second harmonic are also displayed as a green and a cyan fringe, respectively. Another LFQPO is present in the phase 0.84--0.02 as a cyan fringe. We show the phase dependence of the fundamental LFQPO (black points) and the LFQPO in the phase 0.84--0.02 (blue points) in Figure \ref{fig:Rho2QPO}. We show the double of the frequencies of the LFQPO in the phase 0.84--0.02 as red points in panel (a). In the phase 0.84--0.92, the frequency of the fundamental LFQPO is about double of the frequency of the other LFQPO. Considering that the LFQPO with lower frequency evolves smoothly to the phase 0.92--0.02, we argue that the LFQPO in the phase 0.84--0.02 is the sub-harmonic of the fundamental LFQPO in the phase 0.12--0.92. As $\phi$ increases from $\sim 0.18$ to $\sim 0.80$, $f_{\rm tr2}$ decreases from $\sim 5.5$ Hz to $\sim 3.0$ Hz. At $\phi \sim 0.17$ and $\sim 0.85$, $f_{\rm tr2}$ decreases to $\sim 1.5$ Hz. $f_{\rm tr1}$ is not obvious here. In the phase $\sim$ 0.90--0.15 excluding $\sim$ 0.04--0.08 the power density of the aperiodic variability below $\sim 1.5$ Hz is enhanced. This result is further demonstrated by the amplitude of the 0.5--1.0 Hz aperiodic variability as a function of phase presented in panel (c) of Figure 3.

\subsection{Energy-frequency-power maps}
 
Figures \ref{fig:F2E40703} and \ref{fig:F2E60405} show the energy-frequency-power maps. The colour scale is the same as that of the phase-frequency-power maps. The LFQPOs and their harmonics are displayed as vertical pencil-like patterns in the maps. These maps indicate that the power density is larger in the high-energy/low-frequency part of the maps than in the low-energy/high-frequency part for all phases of both $\rho$ classes, and is larger in the high-energy/low-frequency part than in the low-energy/low-frequency part for phases III, IV, V in the $\rho_1$ class and for phases iii, vi, v in the $\rho_2$ class. For other phases, the power density decreases in the high-energy/intermediate-frequency part while increases in the low-energy/low-frequency part.

\subsection{Phase-lag spectra}

\begin{figure*}
\centerline{\includegraphics[height=7.4cm,angle=0]{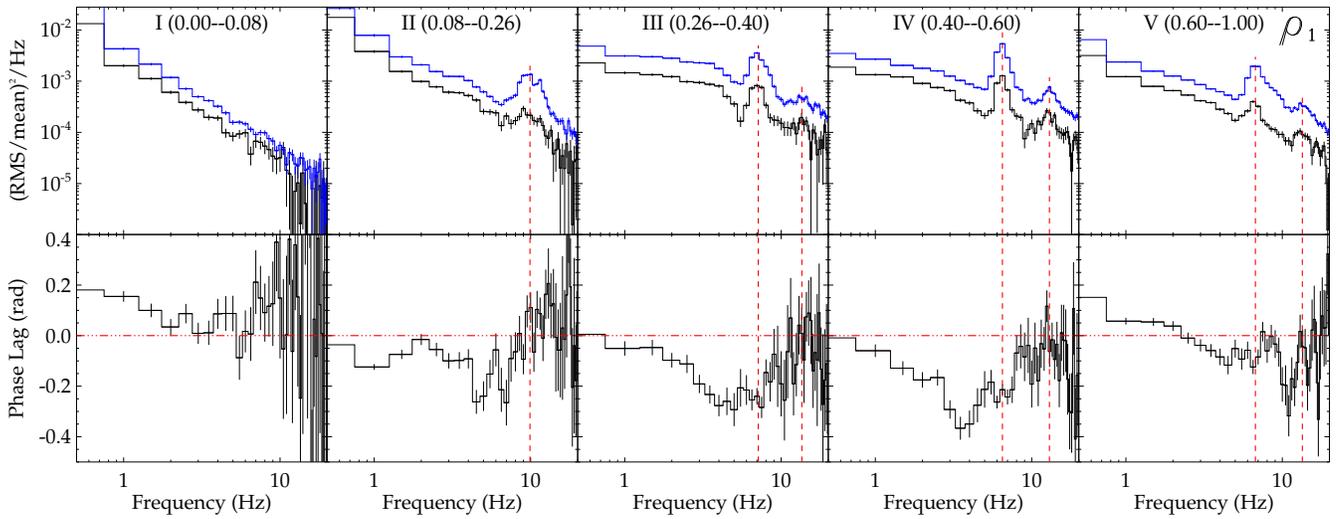}}
\caption{\label{fig:lag40703} The PDS and the phase-lag spectra for different phases of the $\rho_1$ class in GRS 1915+105. The top panels show the PDS in the soft band (1.9--5.1 keV; black lines) and in the hard band (5.1--38.4 keV; blue lines). In the bottom panels a positive lag means that the hard band photons lag the soft band photons. The vertical red dashed lines denote the centroid frequencies of the LFQPO and its harmonic. The horizontal red dot-dashed line denotes the zero phase lag.}
\end{figure*}

\begin{figure*}
\centerline{\includegraphics[height=6.1cm,angle=0]{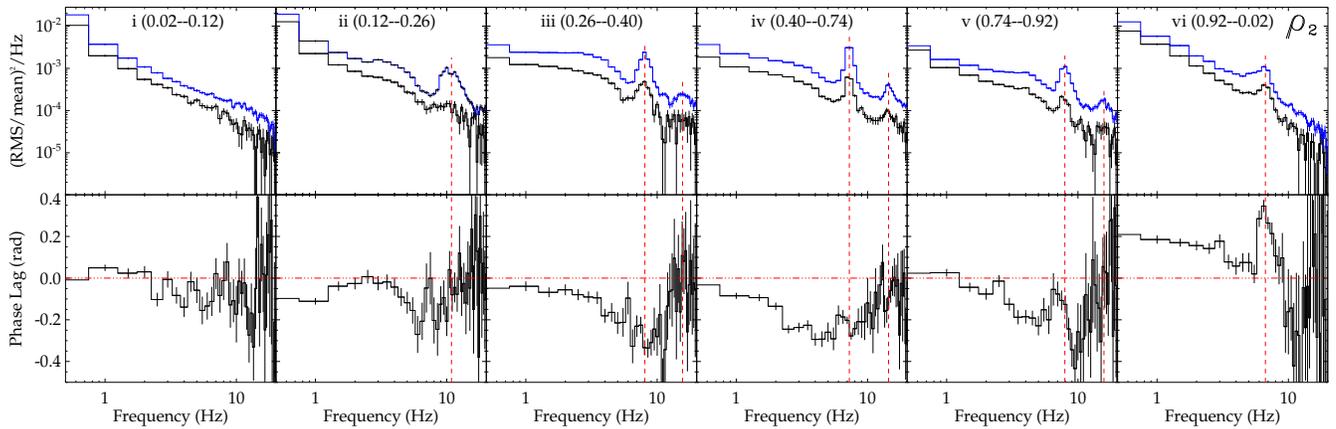}}
\caption{\label{fig:lag60405} Same as Figure \ref{fig:lag40703} for different phases of the $\rho_2$ class in GRS 1915+105. Here the soft band is 2.1--4.9 keV, and the hard band is 4.9--37.8 keV.}
\end{figure*}

Figures \ref{fig:lag40703} and \ref{fig:lag60405} show the PDS and the phase-lag spectra in different phases of the $\rho_1$ and $\rho_2$ classes, respectively. Though the details are different, the phase-lag spectra of the aperiodic variabilities of the two $\rho$ classes have a common feature: the phase lag is usually lower in the intermediate-frequency band than in the low-frequency band. When the source is in the rise phase of the pulse (e.g. phases V in the $\rho_1$ class and vi in the $\rho_2$ class), the phase lag is hard in the low-frequency band (the corresponding time lag, $\tau=\varphi/2 \pi f$, where $f$ is Fourier frequency and $\varphi$ is phase lag, is up to $\sim $ 50 ms). In the high-frequency band the phase lag approximates zero for the $\rho_2$ class and is soft for a narrow range of frequencies for the $\rho_1$ class. When the source is in the slow rise phase (e.g. phases III, IV in the $\rho_1$ class and iii, iv in the $\rho_2$ class), the phase lag is soft ($\tau$ is up to $\sim $ 20 ms). A main difference between the two $\rho$ classes is that at frequencies below $\sim $ 3 Hz the phase lag approximates zero in phase i of the $\rho_2$ class while it is hard in phase I of the $\rho_1$ class.

For phases III, IV, V of the $\rho_1$ class and phases iii, iv of the $\rho_2$ class there is a dip at LFQPO frequency in the phase-lag curve, indicating that the phase lag of the LFQPO is soft ($\tau$ is about several ms). The phase lag approximates zero for the LFQPO in phase II of the $\rho_1$ class and phase ii of the $\rho_2$ class, and is hard ($\tau$ is about 10 ms) for the LFQPO in phase vi of the $\rho_2$ class. It is soft ($\tau$ is about several ms) for the second harmonic in phase iv of the $\rho_2$ class, while approximates zero for the second harmonic in other phases.

\section{DISCUSSION}\label{sec:discuss}

In this section we discuss the timing results of two \emph{RXTE} observations of GRS~1915+105 during the heartbeat state to investigate the origin of the LFQPO and the aperiodic X-ray variability, and combine the timing results with the spectral results of \citet{Neilsen11,Neilsen12} to obtain a spectral-timing unified picture of the accretion in the heartbeat state of GRS 1915+105. We first introduce some interpretations for the phase lag in order to facilitate the discussion.

In black hole X-ray binaries, the phase/time lag between the high- and low- energy bands in the frequency range $\sim$ 0.1--10 Hz has been studied by many authors as an important property of the X-ray variability \citep[e.g.][]{Morgan97,Cui97,Hua97,Poutanen99,Nowak99,Wijnands99,Cui00,Reig00,Lin00,Kotov2001,Poutanen01,Tomsick01,Arevalo2006,Qu10,Uttley11,Shaposhnikov12,Pahari13,Misra13,Altamirano15,Eijnden2016}. The observed phase lag has been interpreted from several aspects, such as radiation propagation, mass transfer, and property of variability. The hard and soft phase lags of the order of light-travel time could be produced by the Comptonization of the soft photons from the disk by the hot electron in the corona \citep[e.g.][]{Nowak99,Reig00} and the reflection of the Comptonized photons from the disk \citep[e.g.][]{Uttley11,Tripathi11,Wilkins13}. The hard and soft lags of the order of a second could be caused by the differential responses of the inner disk radius and the corona to the mass accretion rate \citep{Mir2016} or by the differential QPO frequency in the soft and hard energy bands \citep{Eijnden2016}. The hard lag of the order of a second could also be produced by the inward propagation of mass accretion fluctuations through the disk \citep[e.g.][]{Lyubarskii97,Kotov2001,Arevalo2006}.

\subsection{LFQPO}\label{sec:discuss1}

For the $\rho_2$ class, the fundamental QPO is present in the phase 0.12--0.92, while its sub-harmonic is present in the phase 0.84--0.02 (Figures \ref{fig:f2t} and \ref{fig:Rho2QPO}), indicating that the sub-harmonic QPO can still exist when the fundamental QPO disappears.

We use the model of \citet{Eijnden2016} and the model of \citet{Mir2016} to discuss the lag of the QPO since these two models can explain both hard and soft lags of the order of few tens of milliseconds (which is the case here).

For the first time \citet{Eijnden2016} showed that the energy band with higher QPO frequency was running away from the other band on short timescales, namely, a positive correlation between QPO frequency and photon energy would result in a soft lag, while an anti-correlation between QPO frequency and photon energy would result in a hard lag. The LFQPO in phase iv of the $\rho_2$ class can thus be qualitatively explained using this model as the QPO frequency is positively correlated with photon energy \citep{Yan13apj} and the phase lag is soft (Figure \ref{fig:lag60405}).

The model of \citet{Mir2016} is used to interpret the phase lag of a QPO resulted by the oscillation of the inner disk radius, and thus cannot directly explain the QPOs studied here which are highly correlated with the corona \citep{Yan13apj}. Nevertheless, its concept deserves consideration. The basic idea of the model is that the differential responses of different parts of the accretion flow with different temperatures to the mass accretion rate produce the phase lag. If the mass accretion rate has a QPO variation and the different parts of the corona located above different disk radii are different in temperature, then a hard lag will be produced when the temperature of the corona is larger at smaller radii than at larger radii, otherwise a soft lag will be produced. However, it is not clear whether this model can explain the observed energy dependence of QPO frequency.

\subsection{Aperiodic X-ray variability}

The phase-frequency-power maps show that the power density of the intermediate-frequency aperiodic X-ray variability decreases as the source becomes soft in the slow rise phase (Figure \ref{fig:f2t}). The significant enhancement of the power density of the low-frequency aperiodic variability in the phase $\sim$ 0.92--0.15 of $\rho_1$ class and in the phase $\sim$ 0.90--0.15 excluding $\sim$ 0.04--0.08 of the $\rho_2$ class coincides with the minimum inner radius of the accretion disk \citep[see][]{Neilsen12}, suggesting that at this time part of the low-frequency aperiodic variability is connected with the disk. However, the power density of the low-frequency aperiodic variability in the phase $\sim$ 0.04--0.08 of the $\rho_2$ class is not enhanced. The phase $\sim$ 0.04--0.08 is included in the phase of the second, hard peak of the phase-folded light curve, which is suggested to be the result from the collision between the corona and the material ejected from the inner disk \citep{Neilsen11}. The suppression of the power density of the low-frequency aperiodic variability in the phase $\sim$ 0.04--0.08, which mimics the variability suppression in the phase of the hard peak of the $\rho_2$ class \citep[see Figure 4 in][]{Neilsen11}, is thus possibly due to the ejection of the inner disk. 

The gradient of the power density in the energy-frequency-power maps clearly shows that most of the aperiodic X-ray variability comes from the corona in the slow rise phase (Figures \ref{fig:F2E40703} and \ref{fig:F2E60405}). The energy-frequency-power maps also indicate that in the slow rise phase the low-frequency aperiodic variability from the corona is significant while in the soft $\rho$-cycle phase the low-frequency aperiodic variabilities from the corona and the disk are both significant, consistent with the results of the phase-frequency-power maps. These results also demonstrate that the energy-frequency-power map is an effective new method for studying the origin of the X-ray variability.

A natural interpretation of the large ($\sim$ 50 ms) hard lag at low-frequencies in the soft $\rho$-cycle phases (Figures \ref{fig:lag40703} and \ref{fig:lag60405}) is that the mass accretion fluctuations in the disk propagate inward and drive the corona variability at smaller radii. The millisecond level phase lag at high-frequency may be due to Comptonization and/or reverberation processes. The observed large (up to $\sim$ 20 ms) soft lag cannot be interpreted as a simple reflection delay which is several milliseconds here. The large soft lag could be explained with the scenario of \citet{Mir2016}. It also could be caused by propagation of acoustic waves from the hotter corona toward the cooler disk region. In section \ref{sec:st}, we continue to interpret the phase lag in the context of a spectral-timing combination analysis.

\subsection{Spectral-timing unified picture}\label{sec:st}

In the slow rise phase, the aperiodic X-ray variability from the corona dominates the X-ray variability (Figures \ref{fig:F2E40703} and \ref{fig:F2E60405}) and the phase lag is soft (Figures \ref{fig:lag40703} and \ref{fig:lag60405}), indicating that the X-ray variability is produced in the corona initiatively and then induces the aperiodic variability from the disk. The spectral analysis by \citet{Neilsen11,Neilsen12} indicated that in this phase the disk was a local Eddington limit disk that had a critical radius inside of which the disk was in a critical state, where some excess mass was expelled in the form of wind/outflow and the accretion rate was kept to be at the critical rate \citep{Fukue2004}. It is plausible that in this phase the corona inside of the critical radius has a strong variability and affects the disk through wind/outflow and radiation.

In the rise phase of the pulse, the low-frequency aperiodic X-ray variabilities from both the disk and the corona dominate the X-ray variability (Figures \ref{fig:F2E40703} and \ref{fig:F2E60405}) and the phase lag is hard in the low-frequency band (Figures \ref{fig:lag40703} and \ref{fig:lag60405}), indicating that the low-frequency aperiodic variability is produced in the disk initiatively and then drives the corona. Neilsen et al. (2011, 2012) argued that in this phase the disk became unstable due to a thermal-viscous radiation pressure instability and collapsed inward. It is therefore plausible that the variability of the disk is strong and initiative in this phase.

In the hard pulse of the $\rho_2$ class, the variability shows no significant lag (Figure \ref{fig:lag60405}). \citet{Neilsen11} showed that in this phase a warm ($\sim$ 6 keV) optically thick corona, which might be formed from the collision of the hot corona and the cold material ejected from the inner disk, scattered almost all of the photons from the disk. This process will blur the phase lag between the warm corona and the disk and results in the observed zero phase lag. For the decay phase of the pulse of the $\rho_1$ class, the phase lag below $\sim$ 3 Hz is significant, which is consistent with the spectral result that about 10$\%$ of the disk photons have not been scattered \citep{Neilsen12}.

In short, we obtained a spectral-timing unified picture: when the disk is in a local Eddington limit, inside of the critical radius part of the mass is expelled by radiation pressure, and the aperiodic variability from the corona is initiative and drives the aperiodic variability from the disk; when there is a radiation pressure instability in the disk, the low-frequency aperiodic variability is initiative and drives the low-frequency aperiodic variability from the corona. When the disk photons are completely scattered by the optically thick corona, no significant phase lag has been observed.

\section{CONCLUSIONS}\label{sec:con}

We have performed a detailed timing analysis and made a spectral-timing combination analysis for two {\it RXTE} observations of GRS 1915+105 during the single-peaked heartbeat state ($\rho_1$ class) and the double-peaked heartbeat state ($\rho_2$ class), respectively.

The phase-frequency-power maps indicate that in the slow rise phase the intermediate-frequency aperiodic X-ray variability weakens as the source softens and in the rise phase of the pulse when the disk inner radius decreases the low-frequency aperiodic variability becomes more significant, and for the $\rho_2$ class the LFQPO disappears in the rise phase of the pulse while its sub-harmonic is still present with a hard phase lag.

In the slow rise phase, the energy-frequency-power maps indicate that most of the aperiodic X-ray variability is from the corona, the phase-lag spectra indicate that the phase lag is soft at low- and intermediate-frequencies with a time delay up to 20 ms. In the rise phase of the pulse, the low-frequency aperiodic variabilities from the corona and the disk are both significant, the phase lag is hard at low-frequencies with a time delay up to 50 ms. In the hard pulse of the $\rho_2$ class the phase lag approximates zero.

A spectral-timing unified picture is derived from the combination of our timing results and the spectral results of \citet{Neilsen11,Neilsen12}. When the disk is in a local Eddington limit the aperiodic variability from the corona drives the aperiodic variability from the disk, when the disk is in a radiation pressure instability the low-frequency aperiodic variability from the disk drives the low-frequency aperiodic variability from the corona. In the hard pulse of the $\rho_2$ class the zero phase lag may be resulted from a fully scattering of the disk photons by the optically thick corona.

\section*{Acknowledgements}
We thank the anonymous reviewer for comments which greatly improved the quality and clarity of the paper. We thank Kinwah Wu, Peng-Fei Chen and Min Long for helpful discussions. The research has made use of data obtained from the High Energy Astrophysics Science Archive Research Center (HEASARC), provided by NASA's Goddard Space Flight Center. This work is supported by National Natural Science Foundation of China (grant Nos. 11273062, 11133001, 11333004, 11173041 and 11373006), the Strategic Priority Research Program of CAS (grant no. XDB09000000), and China Postdoctoral Science Foundation (grant No. 2015M571838). Li Ji is also supported by the 100 Talents program of Chinese Academy of Sciences.

\bibliographystyle{mn2e}
\bibliography{yan}

\begin{thebibliography}{57}
\expandafter\ifx\csname natexlab\endcsname\relax\def\natexlab#1{#1}\fi

\bibitem[{{Altamirano} \& {M{\'e}ndez}(2015)}]{Altamirano15}
{Altamirano}, D., \& {M{\'e}ndez}, M. 2015, \mnras, 449, 4027

\bibitem[{{Ar{\'e}valo} \& {Uttley}(2006)}]{Arevalo2006}
{Ar{\'e}valo}, P., \& {Uttley}, P. 2006, \mnras, 367, 801

\bibitem[{{Belloni} {et~al.}(2000){Belloni}, {Klein-Wolt}, {M{\'e}ndez}, {van
  der Klis}, \& {van Paradijs}}]{Belloni00}
{Belloni}, T., {Klein-Wolt}, M., {M{\'e}ndez}, M., {van der Klis}, M., \& {van
  Paradijs}, J. 2000, \aap, 355, 271

\bibitem[{{Belloni} {et~al.}(2001){Belloni}, {M{\'e}ndez}, \&
  {S{\'a}nchez-Fern{\'a}ndez}}]{Belloni01}
{Belloni}, T., {M{\'e}ndez}, M., \& {S{\'a}nchez-Fern{\'a}ndez}, C. 2001, \aap,
  372, 551

\bibitem[{{Belloni} {et~al.}(2002){Belloni}, {Psaltis}, \& {van der
  Klis}}]{Belloni02}
{Belloni}, T., {Psaltis}, D., \& {van der Klis}, M. 2002, \apj, 572, 392

\bibitem[{{Belloni} {et~al.}(2006){Belloni}, {Soleri}, {Casella}, {M{\'e}ndez},
  \& {Migliari}}]{Belloni06}
{Belloni}, T., {Soleri}, P., {Casella}, P., {M{\'e}ndez}, M., \& {Migliari}, S.
  2006, \mnras, 369, 305

\bibitem[{{Berger} \& {van der Klis}(1994)}]{Berger94}
{Berger}, M., \& {van der Klis}, M. 1994, \aap, 292, 175

\bibitem[{{Castro-Tirado} {et~al.}(1992){Castro-Tirado}, {Brandt}, \&
  {Lund}}]{Castro92}
{Castro-Tirado}, A.~J., {Brandt}, S., \& {Lund}, N. 1992, \iaucirc, 5590, 2

\bibitem[{{Chen} {et~al.}(1997){Chen}, {Swank}, \& {Taam}}]{Chen97}
{Chen}, X., {Swank}, J.~H., \& {Taam}, R.~E. 1997, \apjl, 477, L41

\bibitem[{{Cui} {et~al.}(2000){Cui}, {Zhang}, \& {Chen}}]{Cui00}
{Cui}, W., {Zhang}, S.~N., \& {Chen}, W. 2000, \apjl, 531, L45

\bibitem[{{Cui} {et~al.}(1997){Cui}, {Zhang}, {Focke}, \& {Swank}}]{Cui97}
{Cui}, W., {Zhang}, S.~N., {Focke}, W., \& {Swank}, J.~H. 1997, \apj, 484, 383

\bibitem[{{Fender} {et~al.}(1999){Fender}, {Garrington}, {McKay}, {Muxlow},
  {Pooley}, {Spencer}, {Stirling}, \& {Waltman}}]{Fender99}
{Fender}, R.~P., {Garrington}, S.~T., {McKay}, D.~J., {Muxlow}, T.~W.~B.,
  {Pooley}, G.~G., {Spencer}, R.~E., {Stirling}, A.~M., \& {Waltman}, E.~B.
  1999, \mnras, 304, 865

\bibitem[{{Fukue}(2004)}]{Fukue2004}
{Fukue}, J. 2004, \pasj, 56, 569

\bibitem[{{Greiner} {et~al.}(2001){Greiner}, {Cuby}, {McCaughrean},
  {Castro-Tirado}, \& {Mennickent}}]{Greiner01b}
{Greiner}, J., {Cuby}, J.~G., {McCaughrean}, M.~J., {Castro-Tirado}, A.~J., \&
  {Mennickent}, R.~E. 2001, \aap, 373, L37

\bibitem[{{Hannikainen} {et~al.}(2005){Hannikainen}, {Rodriguez}, \& {et
  al.}}]{Hannikainen05}
{Hannikainen}, D.~C., {Rodriguez}, J., \& {et al.}, V. 2005, \aap, 435, 995

\bibitem[{{Harlaftis} \& {Greiner}(2004)}]{Harlaftis04}
{Harlaftis}, E.~T., \& {Greiner}, J. 2004, \aap, 414, L13

\bibitem[{{Hua} {et~al.}(1997){Hua}, {Kazanas}, \& {Titarchuk}}]{Hua97}
{Hua}, X.-M., {Kazanas}, D., \& {Titarchuk}, L. 1997, \apjl, 482, L57

\bibitem[{{Klein-Wolt} {et~al.}(2002){Klein-Wolt}, {Fender}, {Pooley},
  {Belloni}, {Migliari}, {Morgan}, \& {van der Klis}}]{Klein02}
{Klein-Wolt}, M., {Fender}, R.~P., {Pooley}, G.~G., {Belloni}, T., {Migliari},
  S., {Morgan}, E.~H., \& {van der Klis}, M. 2002, \mnras, 331, 745

\bibitem[{{Kotov} {et~al.}(2001){Kotov}, {Churazov}, \& {Gilfanov}}]{Kotov2001}
{Kotov}, O., {Churazov}, E., \& {Gilfanov}, M. 2001, \mnras, 327, 799

\bibitem[{{Lin} {et~al.}(2000){Lin}, {Smith}, {Liang}, {Bridgman}, {Smith},
  {Mart{\'{\i}}}, {Durouchoux}, {Mirabel}, \& {Rodr{\'{\i}}guez}}]{Lin00}
{Lin}, D., {et~al.} 2000, \apj, 532, 548

\bibitem[{{Lyubarskii}(1997)}]{Lyubarskii97}
{Lyubarskii}, Y.~E. 1997, \mnras, 292, 679

\bibitem[{{Massaro} {et~al.}(2010){Massaro}, {Ventura}, {Massa}, {Feroci},
  {Mineo}, {Cusumano}, {Casella}, \& {Belloni}}]{Massaro10}
{Massaro}, E., {Ventura}, G., {Massa}, F., {Feroci}, M., {Mineo}, T.,
  {Cusumano}, G., {Casella}, P., \& {Belloni}, T. 2010, \aap, 513, A21

\bibitem[{{McClintock} {et~al.}(2006){McClintock}, {Shafee}, {Narayan},
  {Remillard}, {Davis}, \& {Li}}]{McClintock06}
{McClintock}, J.~E., {Shafee}, R., {Narayan}, R., {Remillard}, R.~A., {Davis},
  S.~W., \& {Li}, L.-X. 2006, \apj, 652, 518

\bibitem[{{Middleton} {et~al.}(2006){Middleton}, {Done}, {Gierli{\'n}ski}, \&
  {Davis}}]{Middleton06}
{Middleton}, M., {Done}, C., {Gierli{\'n}ski}, M., \& {Davis}, S.~W. 2006,
  \mnras, 373, 1004

\bibitem[{{Miller} {et~al.}(2014){Miller}, {Mineshige}, {Kubota}, \& {the
  ASTRO-H Science Working Group}}]{Miller14b}
{Miller}, J.~M., {Mineshige}, S., {Kubota}, A., \& {the ASTRO-H Science Working
  Group}. 2014, ArXiv:1412.1173

\bibitem[{{Miller} {et~al.}(2013){Miller}, {Parker}, {Fuerst}, {Bachetti},
  {Harrison}, {Barret}, {Boggs}, {Chakrabarty}, {Christensen}, {Craig},
  {Fabian}, {Grefenstette}, {Hailey}, {King}, {Stern}, {Tomsick}, {Walton}, \&
  {Zhang}}]{Miller13}
{Miller}, J.~M., {et~al.} 2013, \apjl, 775, L45

\bibitem[{{Mir} {et~al.}(2016){Mir}, {Misra}, {Pahari}, {Iqbal}, \&
  {Ahmad}}]{Mir2016}
{Mir}, M.~H., {Misra}, R., {Pahari}, M., {Iqbal}, N., \& {Ahmad}, N. 2016,
  \mnras, 457, 2999

\bibitem[{{Mirabel} \& {Rodr{\'{\i}}guez}(1994)}]{Mirabel94}
{Mirabel}, I.~F., \& {Rodr{\'{\i}}guez}, L.~F. 1994, \nat, 371, 46

\bibitem[{{Misra} \& {Mandal}(2013)}]{Misra13}
{Misra}, R., \& {Mandal}, S. 2013, \apj, 779, 71

\bibitem[{{Miyamoto} {et~al.}(1992){Miyamoto}, {Kitamoto}, {Iga}, {Negoro}, \&
  {Terada}}]{Miyamoto92}
{Miyamoto}, S., {Kitamoto}, S., {Iga}, S., {Negoro}, H., \& {Terada}, K. 1992,
  \apjl, 391, L21

\bibitem[{{Morgan} {et~al.}(1997){Morgan}, {Remillard}, \&
  {Greiner}}]{Morgan97}
{Morgan}, E.~H., {Remillard}, R.~A., \& {Greiner}, J. 1997, \apj, 482, 993

\bibitem[{{Neilsen} {et~al.}(2011){Neilsen}, {Remillard}, \& {Lee}}]{Neilsen11}
{Neilsen}, J., {Remillard}, R.~A., \& {Lee}, J.~C. 2011, \apj, 737, 69

\bibitem[{{Neilsen} {et~al.}(2012){Neilsen}, {Remillard}, \& {Lee}}]{Neilsen12}
---. 2012, \apj, 750, 71

\bibitem[{{Nowak}(2000)}]{Nowak00}
{Nowak}, M.~A. 2000, \mnras, 318, 361

\bibitem[{{Nowak} {et~al.}(1999){Nowak}, {Vaughan}, {Wilms}, {Dove}, \&
  {Begelman}}]{Nowak99}
{Nowak}, M.~A., {Vaughan}, B.~A., {Wilms}, J., {Dove}, J.~B., \& {Begelman},
  M.~C. 1999, \apj, 510, 874

\bibitem[{{Pahari} {et~al.}(2013){Pahari}, {Neilsen}, {Yadav}, {Misra}, \&
  {Uttley}}]{Pahari13}
{Pahari}, M., {Neilsen}, J., {Yadav}, J.~S., {Misra}, R., \& {Uttley}, P. 2013,
  \apj, 778, 136

\bibitem[{{Poutanen}(2001)}]{Poutanen01}
{Poutanen}, J. 2001, Advances in Space Research, 28, 267

\bibitem[{{Poutanen} \& {Fabian}(1999)}]{Poutanen99}
{Poutanen}, J., \& {Fabian}, A.~C. 1999, \mnras, 306, L31

\bibitem[{{Qu} {et~al.}(2010){Qu}, {Lu}, {Lu}, {Song}, {Zhang}, {Ding}, \&
  {Wang}}]{Qu10}
{Qu}, J.~L., {Lu}, F.~J., {Lu}, Y., {Song}, L.~M., {Zhang}, S., {Ding}, G.~Q.,
  \& {Wang}, J.~M. 2010, \apj, 710, 836

\bibitem[{{Reid} {et~al.}(2014){Reid}, {McClintock}, {Steiner}, {Steeghs},
  {Remillard}, {Dhawan}, \& {Narayan}}]{Reid14}
{Reid}, M.~J., {McClintock}, J.~E., {Steiner}, J.~F., {Steeghs}, D.,
  {Remillard}, R.~A., {Dhawan}, V., \& {Narayan}, R. 2014, \apj, 796, 2

\bibitem[{{Reig} {et~al.}(2000){Reig}, {Belloni}, {van der Klis}, {M{\'e}ndez},
  {Kylafis}, \& {Ford}}]{Reig00}
{Reig}, P., {Belloni}, T., {van der Klis}, M., {M{\'e}ndez}, M., {Kylafis},
  N.~D., \& {Ford}, E.~C. 2000, \apj, 541, 883

\bibitem[{{Rodriguez} \& {Varni{\`e}re}(2011)}]{Rodriguez11}
{Rodriguez}, J., \& {Varni{\`e}re}, P. 2011, \apj, 735, 79

\bibitem[{{Shaposhnikov}(2012)}]{Shaposhnikov12}
{Shaposhnikov}, N. 2012, \apjl, 752, L25

\bibitem[{{Strohmayer}(2001)}]{Strohmayer01}
{Strohmayer}, T.~E. 2001, \apjl, 554, L169

\bibitem[{{Strohmayer} \& {Watts}(2005)}]{Strohmayer2005}
{Strohmayer}, T.~E., \& {Watts}, A.~L. 2005, \apjl, 632, L111

\bibitem[{{Taam} {et~al.}(1997){Taam}, {Chen}, \& {Swank}}]{Taam97}
{Taam}, R.~E., {Chen}, X., \& {Swank}, J.~H. 1997, \apjl, 485, L83

\bibitem[{{Tomsick} \& {Kaaret}(2001)}]{Tomsick01}
{Tomsick}, J.~A., \& {Kaaret}, P. 2001, \apj, 548, 401

\bibitem[{{Tripathi} {et~al.}(2011){Tripathi}, {Misra}, {Dewangan}, \&
  {Rastogi}}]{Tripathi11}
{Tripathi}, S., {Misra}, R., {Dewangan}, G., \& {Rastogi}, S. 2011, \apjl, 736,
  L37

\bibitem[{{Uttley} {et~al.}(2011){Uttley}, {Wilkinson}, {Cassatella}, {Wilms},
  {Pottschmidt}, {Hanke}, \& {B{\"o}ck}}]{Uttley11}
{Uttley}, P., {Wilkinson}, T., {Cassatella}, P., {Wilms}, J., {Pottschmidt},
  K., {Hanke}, M., \& {B{\"o}ck}, M. 2011, \mnras, 414, L60

\bibitem[{{van den Eijnden} {et~al.}(2016){van den Eijnden}, {Ingram}, \&
  {Uttley}}]{Eijnden2016}
{van den Eijnden}, J., {Ingram}, A., \& {Uttley}, P. 2016, \mnras, 458, 3655

\bibitem[{{Vilhu} \& {Nevalainen}(1998)}]{Vilhu98}
{Vilhu}, O., \& {Nevalainen}, J. 1998, \apjl, 508, L85

\bibitem[{{Wijnands} {et~al.}(1999){Wijnands}, {Homan}, \& {van der
  Klis}}]{Wijnands99}
{Wijnands}, R., {Homan}, J., \& {van der Klis}, M. 1999, \apjl, 526, L33

\bibitem[{{Wilkins} \& {Fabian}(2013)}]{Wilkins13}
{Wilkins}, D.~R., \& {Fabian}, A.~C. 2013, \mnras, 430, 247

\bibitem[{{Yan} {et~al.}(2013{\natexlab{a}}){Yan}, {Ding}, {Wang}, {Qu}, \&
  {Song}}]{Yan13mn}
{Yan}, S.-P., {Ding}, G.-Q., {Wang}, N., {Qu}, J.-L., \& {Song}, L.-M.
  2013{\natexlab{a}}, \mnras, 434, 59

\bibitem[{{Yan} {et~al.}(2013{\natexlab{b}}){Yan}, {Wang}, {Ding}, \&
  {Qu}}]{Yan13apj}
{Yan}, S.-P., {Wang}, N., {Ding}, G.-Q., \& {Qu}, J.-L. 2013{\natexlab{b}},
  \apj, 767, 44

\bibitem[{{Zdziarski} {et~al.}(2005){Zdziarski}, {Gierli{\'n}ski}, {Rao},
  {Vadawale}, \& {Miko{\l}ajewska}}]{Zdziarski05}
{Zdziarski}, A.~A., {Gierli{\'n}ski}, M., {Rao}, A.~R., {Vadawale}, S.~V., \&
  {Miko{\l}ajewska}, J. 2005, \mnras, 360, 825

\bibitem[{{Zhang} {et~al.}(1997){Zhang}, {Cui}, \& {Chen}}]{Zhang97}
{Zhang}, S.~N., {Cui}, W., \& {Chen}, W. 1997, \apjl, 482, L155

\end{thebibliography}

\label{lastpage}
\end{document}